\documentclass[prl,aps,twocolumn,preprintnumbers,showpacs]{revtex4-1}
\usepackage{amsmath}
\usepackage{amssymb}
\usepackage{graphicx}
\usepackage{mathrsfs}

\newcommand{\Tr}{\mathrm{Tr}}

\begin{document}
\title{Non-Markovian quantum dissipative processes with the same positive features as Markovian dissipative processes}
\author{Da-Jian Zhang$^{1,2}$}
\author{Hua-Lin Huang$^2$}
\author{D. M. Tong$^1$}
\email{tdm@sdu.edu.cn}
\affiliation{$^1$Department of Physics, Shandong University, Jinan 250100, China\\
$^2$School of Mathematics, Shandong University, Jinan 250100, China}
\date{\today}
\begin{abstract}
It has been found that Markovian quantum dissipative processes, described by the Lindblad equation, may have attractive steady-state manifolds, in which dissipation and decoherence can play a positive role to quantum information processing. In this article, we show that such attractive steady-state manifolds with the same positive features as in Markovian dissipative processes can exist in the non-Markovian dissipative processes. Our finding indicates that the dissipation-assisted schemes implemented in the Markovian systems can be directly generalized to the non-Markovian systems.
\end{abstract}
\maketitle
Any real quantum system inevitably interacts with its environment, and hence quantum decoherence is an ubiquitous phenomenon. Yet, decoherence may collapse the desired coherence of a quantum state, and reduce the efficiency of quantum information processing. It is one of the main practical obstacles in quantum information processing. Interestingly, it was recently found that dissipation and decoherence may even play a positive role in quantum information processing. A substantial number of dissipation-assisted schemes have been proposed for various aims of quantum information processing, such as quantum state engineering \cite{Poyatos1996,Kraus2008,Diehl2008,Kastoryano2011,Vollbrecht2011,Krauter2011,Torre2013}, quantum simulation \cite{Barreiro2000,Barreiro2011,Zoller2014}, and quantum computation \cite{Verstraete2009,Zanardi2014}, in which dissipation is no longer undesirable but plays an integral part.

The basic idea of dissipation-assisted schemes is to utilize both the coherence stability of steady states and the effects of dissipation to protect quantum information from decoherence. Information is encoded in the steady-state manifold that possesses some computationally desirable properties, e.g., decoherence-free subspaces \cite{DFS-NS,DFS-NS2} or  noiseless subsystems \cite{DFS-NS3,DFS-NS4}, over which the dynamics is unitary. Dissipation is then utilized to drive the system to the steady-state manifold and suppress the leakage outside the manifold. Implementation of dissipation-assisted schemes is to resort to Markovian dissipative processes, described by the Lindblad master equation,
\begin{eqnarray}
\frac{d}{dt}\rho(t)=\mathcal{L}\rho(t),\label{Lindblad-eq}
\end{eqnarray}
where $\rho(t)$ is a density operator, and $\mathcal{L}$ is a time-independent generator, defined as
$\mathcal{L}\rho=-i[H,\rho]+\sum_\alpha \gamma_\alpha \left[A_\alpha\rho A_\alpha^\dagger-\frac{1}{2}\{A_\alpha^\dagger A_\alpha,\rho\}\right]$  \cite{Lindblad}. Here, $A_\alpha$ are time-independent Lindblad operators,  $\gamma_\alpha>0$ are positive decay rates, and $H$ is a time-independent Hamiltonian. Equation (\ref{Lindblad-eq}) can induce a decomposition of the Hilbert space with the form
$\mathcal{H}=\oplus_\alpha\mathcal{H}_{\alpha,1}\otimes\mathcal{H}_{\alpha,2}\oplus\mathcal{K}$,
where $\mathcal{H}_{\alpha,1}$ is a noiseless subsystem, $\mathcal{H}_{\alpha,2}$ is a noiseful subsystem that supports a unique fixed density operator, and $\mathcal{K}$ is a decaying subspace \cite{Baumgartner}. As a consequence of the decomposition, there always exists the steady-state manifold in the Markovian processes, and its structure reads $\rho_{\textrm{ss}}=\sum_\alpha p_\alpha\rho_{\alpha,1}\otimes\rho_{\alpha,2}$,
where $\rho_{\alpha,1}$ is an arbitrary density operator on $\mathcal{H}_{\alpha,1}$, $\rho_{\alpha,2}$ is a unique fixed density operator on $\mathcal{H}_{\alpha,2}$, and $p_\alpha$ is an arbitrary non-negative number subject to $\sum_\alpha p_\alpha=1$. The steady states remain unchanged during the evolution of the system while non-steady states suffer from exponential decay resulting from the effects of dissipation and decoherence, i.e., the steady-state manifold is attractive, provided that the attractive condition, $\min_{\lambda\neq 0}|\textrm{Re}(\lambda)|>0$, is fulfilled,  where $\lambda$ are the eigenvalues of $\mathcal{L}$. The evolution of the open system can be then confined to the steady-state manifold in the long-time limit. By taking advantage of the positive features of the Markovian dissipative processes, i.e., the existence, the structure, and the attractiveness of the steady-state manifold, one can achieve the goal of both maintaining coherence and suppressing leakage, thus implementing dissipation-assisted schemes successfully.

However, the Markovian process described by the Lindblad master equation (\ref{Lindblad-eq}) is only an approximation relying on a number of simplifications, which is usually valid in the weak-coupling limit \cite{Breuer2007}. Real physical systems often feature non-Markovianity and evolve in the processes described by the general master equation, i.e., the time-convolutionless master equation,
\begin{eqnarray}
\frac{d}{dt}\rho(t)=\mathcal{L}(t)\rho(t), \label{TCL-eqn}
\end{eqnarray}
where $\mathcal{L}(t)$ is a time-dependent superoperator known as the time-convolutionless generator \cite{Breuer2007}.
Do any non-Markovian quantum systems have the attractive steady-state manifolds with the same structure as in Markovian dissipative systems? In this article, we address this issue. We will show that such attractive steady-state manifolds exist in the non-Markovian processes described by Eq. (\ref{TCL-eqn}) with the commutative time-dependent generator,
\begin{eqnarray}
[\mathcal{L}(t),\mathcal{L}(t^\prime)]=0 \label{comm-condition},
\end{eqnarray}
such that the dissipation-assisted schemes of Markovian systems can be directly generalized to the non-Markovian systems.

We present our result step by step in the following.

{\sl First}, we prove the existence of the steady-state manifold in the non-Markovian process expressed by Eq. (\ref{TCL-eqn}) with Eq. (\ref{comm-condition}).

The dynamics of the open system can be expressed in terms of a completely positive and trace-preserving (CPTP) map, $\Lambda(t):=\textrm{exp}\left(\int_0^t\mathcal{L}(\tau)d\tau\right)$, transforming the initial state $\rho(0)$ to the state $\rho(t)=\Lambda(t)\rho(0)$ at time $t$. Hereafter, we denote by $\cal{H}$ ($\textrm{dim}(\mathcal{H})<\infty$) the Hilbert space of the system, by
$D(\mathcal{H})$ the set of density operators on the Hilbert space $\cal{H}$, by $F_t$ the set of fixed points of the map $\Lambda(t)$ at time $t$, i.e., $F_t:=\{\rho\in D(\mathcal{H})|\Lambda(t)\rho=\rho\}$, and by $\mathcal{F}$ the collection of the fixed-point sets indexed by $t$, i.e., $\mathcal{F}:=\{F_t\}_t$.

According to Schauder's fixed point theorem that any continuous map on a convex and compact subset of a Hilbert space has a fixed point, $\Lambda(t)$ has a fixed point. Besides, it is a general property of a finite-dimensional topological vector space, known as the Heine-Borel property, that every closed and bounded subset is compact \cite{Kelley}. It follows that $F_t$ is a nonempty compact set.  Since $[\Lambda(t_1),\Lambda(t_2)]=0$ for any two time points $t_1$ and $t_2$, there is $\Lambda(t_1)\Lambda(t_2)\rho=\Lambda(t_2)\Lambda(t_1)\rho=\Lambda(t_2)\rho$  for all $\rho\in F_{t_1}$.
It implies that $\Lambda(t_2)$ maps $F_{t_1}$ into $F_{t_1}$. Using again Schauder's fixed point theorem, we obtain that $\Lambda(t_2)$ has a fixed point belonging to $F_{t_1}$, i.e., inside the fixed-point set of the dynamical map at time $t_1$, one can find a subset consisting of fixed points of the dynamical map at time $t_2$. Thus, $F_{t_1}\cap F_{t_2}\neq\emptyset$.

Since the intersection of two compact and convex sets, $F_{t_1}\cap F_{t_2}$,  is compact and convex, the above procedure can be repeated, e.g., inside $F_{t_1}\cap F_{t_2}$, one can find a subset consisting of fixed points of the dynamical map at time $t_3$, and hence $F_{t_1}\cap F_{t_2}\cap F_{t_3}\neq\emptyset$. Continuing in this manner, we obtain $F_{t_1}\cap F_{t_2}\cap\cdots\cap F_{t_n}\neq\emptyset$ for $n$ being a finite integer. Hence, the collection $\mathcal{F}$ has the finite intersection property \cite{Kelley}. From compactness property of $F_t$ and the finite intersection property of $\mathcal{F}$, it follows that $\bigcap_t F_t\neq\emptyset$ \cite{Kelley}. This completes the proof of the existence of steady states.

{\sl Second}, we analyze the structure of the steady-state manifold existing in the non-Markovian process expressed by Eq. (\ref{TCL-eqn}) with Eq. (\ref{comm-condition}).

To this end, we first establish an auxiliary map $\mathcal{P}_\infty(t):=\lim_{N\rightarrow\infty}\mathcal{P}_N(t)$, where $\mathcal{P}_N(t)=\frac{1}{N}\sum_{n=1}^N\Lambda(t)^n$ are the Ces\`{a}ro means of the sequence $\{\Lambda(t)^n,n=1,2,\dots\}$. For a finite-dimensional open system, the limit of $\mathcal{P}_N(t)$ for $N\rightarrow\infty$ always exists.  $\mathcal{P}_\infty(t)$ is a CPTP map satisfying $\mathcal{P}_\infty(t)\Lambda(t)=\Lambda(t)\mathcal{P}_\infty(t)=\mathcal{P}_\infty^2(t)=\mathcal{P}_\infty(t)$. We then have  $\mathcal{P}_\infty(t)\rho=\rho$ for $\rho\in F_t$ and $\Lambda(t)\mathcal{P}_\infty(t)\rho=\mathcal{P}_\infty(t)\rho$ for $\rho\in D(\cal{H})$, which means that $\mathcal{P}_\infty(t)$ is a projection onto $F_t$, i.e., $F_t=\mathcal{P}_\infty(t)[D(\mathcal{H})]$.

Based on the auxiliary map $\mathcal{P}_\infty(t)$, we then establish another auxiliary map $\mathcal{P}$ being the projection onto the set of steady states. Note that $\{\mathcal{P}_\infty(t)\}_t$ is a family of mutually commutative projections, which can be simultaneously diagonalized. The number of distinguishing elements of $\{\mathcal{P}_\infty(t)\}_t$ is finite \cite{Note4}. We use $\mathcal{P}_\infty(t_1),\mathcal{P}_\infty(t_2),\dots,\mathcal{P}_\infty(t_n)$ to represent these distinguishing elements. Let
$\mathcal{P}:=\mathcal{P}_\infty(t_1)\mathcal{P}_\infty(t_2)\cdots\mathcal{P}_\infty(t_n)$. $\mathcal{P}$ is a CPTP map satisfying the conditions
$\mathcal{P}_\infty(t)\mathcal{P}=\mathcal{P}\mathcal{P}_\infty(t)=\mathcal{P}^2=\mathcal{P}$.
We then have $\cal{P}\rho=\rho$ for $\rho\in\bigcap_t F_t$ and $\mathcal{P}_\infty(t)\mathcal{P}\rho=\mathcal{P}\rho$ for $\rho\in D(\cal{H})$,
which means that $\mathcal{P}$ is a projection onto the set of steady states, i.e., $\bigcap_t F_t=\mathcal{P}[D(\mathcal{H})]$.

With the aid of $\mathcal{P}$, we now identify the structure of the steady states. Note that every density operator $\rho$ has a support, denoted by $P_\rho$, which is the the smallest projection operator satisfying $\Tr(\rho P_\rho)=1$. There exists a density operator $\rho_0\in\bigcap_t F_t$ that satisfies $\Tr(\rho P_{\rho_0})=1$ for all $\rho\in\bigcap_t F_t$, which can be obtained by  convex combination. Further, we decompose the Hilbert space as, $\mathcal{H}=\widetilde{\mathcal{H}}\oplus\mathcal{K}$, with $\widetilde{\cal{H}}:=P_{\rho_0}\cal{H}$ and $\mathcal{K}:=(I-P_{\rho_0})\cal{H}$, where $I$ is the identity operator. All steady states are then supported within the subspace $\widetilde{\cal{H}}$. It can be shown that $\widetilde{\mathcal{H}}$ is invariant under the action of $\mathcal{P}$. Indeed, if $\mathcal{P}^*$ is used to denote the dual map of $\mathcal{P}$, there are $\mathcal{P}^*(P_{\rho_0})\leq \mathcal{P}^*(I)=I$ and thus $P_{\rho_0}-P_{\rho_0}\mathcal{P}^*(P_{\rho_0})P_{\rho_0}\geq 0$. Here, $X\leq Y$ means that $(Y-X)$ is positive semi-definite, and $\mathcal{P}$ and $\mathcal{P}^*$ satisfy the relation $\langle X, \mathcal{P}(Y)\rangle=\langle \mathcal{P}^*(X), Y\rangle$ for all operators $X$ and $Y$, where $\langle X, Y\rangle:=\Tr{X^\dagger Y}$ is the Hilbert-Schmidt inner product.
Besides, the fact that $\rho_0$ is a fixed point of $\mathcal{P}$ implies that $\Tr[\rho_0(P_{\rho_0}-P_{\rho_0}\mathcal{P}^*(P_{\rho_0})P_{\rho_0})]=0$. Since $\rho_0$ is a full-rank density operator on the subspace  $\widetilde{\mathcal{H}}$, there is $P_{\rho_0}\mathcal{P}^*(P_{\rho_0})P_{\rho_0}= P_{\rho_0}$. It leads to $\Tr[P_{\rho_0}\mathcal{P}(\rho)]=1$ for all density operators $\rho$ on $\widetilde{\mathcal{H}}$, which means that $\widetilde{\mathcal{H}}$ is an invariant subspace of $\mathcal{P}$.
Hence, we are able to define a map, $\widetilde{\mathcal{P}}:=\mathcal{P}|_{\widetilde{\cal{H}}}$, whose action is identical to that of $\cal{P}$ on the subspace $\widetilde{\cal{H}}$. The map $\widetilde{\mathcal{P}}$ satisfies all the conditions in Ref. \cite{Lindblad1999}, i.e., it is a CPTP map, a projection, and with a full-rank fixed point. Therefore, the map can be explicitly expressed as
\begin{eqnarray}
\widetilde{\mathcal{P}}\rho=\sum_\alpha \Tr_{\alpha,1}(P_\alpha\rho P_\alpha)\otimes\rho_{\alpha,2}, \label{CP-expression}
\end{eqnarray}
corresponding to a certain decomposition of $\widetilde{\mathcal{H}}$,
$\widetilde{\mathcal{H}}=\oplus_\alpha \mathcal{H}_{\alpha,1}\otimes\mathcal{H}_{\alpha,2}$,
where $P_\alpha$ denotes the orthogonal projector onto the subspace $\mathcal{H}_{\alpha,1}\otimes\mathcal{H}_{\alpha,2}$, and $\rho_{\alpha,2}$ is a fixed density operator on $\mathcal{H}_{\alpha,2}$.

From Eq. (\ref{CP-expression}), we obtain the structure of the steady states,
\begin{eqnarray}
\rho_{\textrm{ss}}=\sum_\alpha p_\alpha \rho_{\alpha,1}\otimes\rho_{\alpha,2},\label{stru.}
\end{eqnarray}
corresponding to the decomposition of the Hilbert space $\mathcal{H}=\oplus_\alpha\mathcal{H}_{\alpha,1}\otimes\mathcal{H}_{\alpha,2}\oplus\mathcal{K}$, where $\rho_{\alpha,1}$ is an arbitrary density operator on $\mathcal{H}_{\alpha,1}$, and $p_\alpha$ is an arbitrary non-negative number satisfying $\sum_\alpha p_\alpha=1$.

{\sl Third}, we discuss the attractiveness of the steady-state manifold existing in the non-Markovian process expressed by Eq. (\ref{TCL-eqn}) with Eq. (\ref{comm-condition}).

It is well-known that in the Markovian case, the attractive condition reads $\min_{\lambda\neq 0}|\textrm{Re}(\lambda)|>0$, which is the essential prerequisite condition for the attractiveness of the steady-state manifold. However, in the non-Markovian case, the gap $\textrm{Re}(\lambda)$ may be zero, since the decay rates may take negative values, which is a sign of the non-Markovian memory effects and reflects a flow of information from the environment back to the open system \cite{Breuer2009}.  Nevertheless, by resorting to the accumulation of dissipation-decoherence effects over time instead of the gap, we find that it is still available to achieve the attractiveness of the steady-state manifold in the non-Markovian regime. We now show this point.

To find the attractive condition for the steady-state manifold in the non-Markovian case, we express $\mathcal{L}(t)$ as the form of the spectral decomposition,
\begin{eqnarray}
\mathcal{L}(t)\rho=\sum_\mu\lambda_\mu(t)R_\mu\Tr(L_\mu^\dagger\rho),\label{damping-basis}
\end{eqnarray}
where $\lambda_\mu(\tau)$ are the spectral parameters, corresponding to the gap in the Markovian case, and $R_\mu$ and $L_\mu$, being time-independent because of Eq. (\ref{comm-condition}), define the damping basis for $\mathcal{L}(t)$, satisfying $\Tr(R_\mu L_\nu^\dagger)=\delta_{\mu\nu}$ \cite{Briegel1993}. Equation (\ref{damping-basis}) implies that the dynamical map can be expressed as
\begin{eqnarray}
\Lambda(t)\rho=\sum_\mu e^{\int_0^t\lambda_\mu(t')dt'}R_\mu\Tr(L_\mu^\dagger\rho).\label{CP-map-basis}
\end{eqnarray}
The spectral parameters $\lambda_\mu(t)$ satisfy $\textrm{Re}\left(\int_0^t\lambda_\mu(t')dt'\right)\leq 0$ due to the CPTP property of $\Lambda(t)$. Besides, by using the existence of steady states, we have that there must exist some $\mu$'s, corresponding to the steady-state manifold, such that $\lambda_\mu(t)=0$ during the whole evolution. From Eq. (\ref{CP-map-basis}), we can then conclude that the steady-state manifold is attractive if and only if
\begin{eqnarray}
\left|\textrm{Re}\left(\int_0^t\lambda_\mu(t')dt'\right)\right|\rightarrow\infty, ~~\textrm{when}~~t\rightarrow\infty,\label{conditon-for-attractiveness}
\end{eqnarray}
for all the other $\mu$'s that are not corresponding to the steady-state manifold. Combining Eqs. (\ref{CP-map-basis}) and (\ref{conditon-for-attractiveness}), we immediately have that the evolution of the system will be confined to the steady-state manifold in the long-time limit, which is the desirable property for implementing dissipation-assisted schemes.
It is obvious that when $\lambda_\mu(t)$ is time-independent, Eq. (\ref{conditon-for-attractiveness}) reduces to $\min_{\lambda\neq 0}|\textrm{Re}(\lambda)|t \rightarrow\infty$ for $t\rightarrow\infty$, i.e.,  $\min_{\lambda\neq 0}|\textrm{Re}(\lambda)|>0$, which is just the attractive condition for the Markovian dissipative processes. In this sense, Eq. (\ref{conditon-for-attractiveness}) can be regarded as a generalization of the attractive condition from Markovian dissipative processes to the non-Markovian case.

So far, we have completed the proof that in the non-Markovian process expressed by Eq. (\ref{TCL-eqn}) with Eq. (\ref{comm-condition}), the steady states always exist, the steady-state manifold is with the structure expressed by Eq. (\ref{stru.}), and the steady-state manifold is attractive if and only if Eq. (\ref{conditon-for-attractiveness}) holds.

After having presented the main result, we would like to discuss some aspects of its application in the following.

First, it is instructive to note that a manifold of steady states always exists in the quantum system described by Eq. (\ref{TCL-eqn}) as long as $\mathcal{L}(t)$ satisfies the commutative relation Eq. (\ref{comm-condition}), while the existence is not true in general if Equation (\ref{comm-condition}) is unsatisfied. We would like to take the model in Refs. \cite{Tu2008,Xiong2012} as an example to illustrate this point. The model is a double quantum dot system coupled to an electron reservoir with the Hamiltonian
 $H=H_S+H_B+H_I$, where $H_S=\sum_{n=1}^2\epsilon a_n^\dagger a_n$, $H_B=\sum_k\varepsilon_k b_k^\dagger b_k$, and $H_I=\sum_{n=1}^2\sum_k e^{i\alpha_n}a_n^\dagger\otimes g_k b_k$+\textrm{H.c.}, corresponding to the system, the reservoir, and the interaction between them, respectively. $a_n^\dagger$ and $a_n$ are creation and annihilation operators for the $n$th energy level of the system with energy $\epsilon$, $b_k^\dagger$ and $b_k$ are creation and annihilation operators for $k$th energy level of the reservoir with energy $\varepsilon_k$, and the coupling strength is denoted by $g_k$ with an explicit phase $\alpha_n$. By eliminating completely all the degrees of freedom of the reservoir, one can obtain an exact non-Markovian master equation for the system, in which the decoherence term reads
 \begin{eqnarray}
\mathcal{L}(t)\rho&=&\kappa(t)(2A\rho A^\dagger-\{A^\dagger A, \rho\})\nonumber\\
&&+\widetilde{\kappa}(t)(2A^\dagger\rho A-\{AA^\dagger, \rho\}),\label{example-1}
\end{eqnarray}
where $A=\frac{1}{\sqrt{2}}\left[a_1+e^{i(\alpha_1-\alpha_2)}a_2\right]$ is an effective fermion operator, and $\kappa(t)$ and $\widetilde{\kappa}(t)$ are the decay rates determined microscopically and nonperturbatively.
In general, the generator expressed in Eq. (\ref{example-1}) does not satisfy Eq. (\ref{comm-condition}), and detailed calculations show that there is not a manifold of steady states in this case. However, the system will possess steady states if Eq. (\ref{comm-condition}) is fulfilled, i.e., at least one of the decay rates vanishes. To achieve this goal, one can apply a large bias to raise (or lower) the Fermi surface of the reservoir such that Eq. (\ref{comm-condition}) is satisfied. Further calculations verify that the steady states can constitute a non-trivial decoherence-free subspace, which shows that one can physically realize the decoherence-free subspace in both weakly and strongly non-Markovian regimes \cite{Xiong2012}.

Second, it is reasonable to believe that the non-Markovian process expressed by Eq. (\ref{TCL-eqn}) with Eq. (\ref{comm-condition}) can provide a promising way for implementing dissipation-assisted schemes. On one hand, our result shows that the non-Markovian process expressed by Eq. (\ref{TCL-eqn}) with Eq. (\ref{comm-condition}) possesses the same positive features as the Markovian dissipative process described by the Lindblad equation, which has found its applications in implementing dissipation-assisted schemes. On the other hand, there are indeed a number of open systems fulfilling Eqs. (\ref{TCL-eqn}) with (\ref{comm-condition}), such as the pure decoherence model \cite{Palma1996,Reina2002,Chruscinski2010}, the damped Jaynes-Cummings model \cite{Breuer2007}, the tight-binding quantum diffusion model \cite{Esposito2005}, the quantum transport model \cite{Steinigeweg2007}, and  the model of spontaneous decay of a two-level system \cite{Breuer2007}.  However, in the practical application, there may still be problems needed to be overcome. For instance,  a system fulfilling Eq. (\ref{TCL-eqn}) with Eq. (\ref{comm-condition}) may lack interests for practical purposes, e.g., its steady-state manifold may be only a one-dimensional subspace and hence not sufficient for the purpose of quantum computation. Such problems may be resolved by resorting to quantum control and environment engineering. We would like to take the decoherence model of two qubits in Ref. \cite{Addis2013} as an example to illustrate this point. Without loss of generality, the master equation describing the dynamics of the two qubits reads
\begin{eqnarray}
\frac{d\rho}{dt}=\frac{\gamma_1(t)-\gamma_2(t)}{2}\mathcal{L}_1\rho+\frac{\gamma_1(t)+\gamma_2(t)}{2}\mathcal{L}_2\rho,\label{example-2}
\end{eqnarray}
with $\mathcal{L}_1\rho:=(\sigma_z^A-\sigma_z^B)\rho(\sigma_z^A-\sigma_z^B)
-\frac{1}{2}\{(\sigma_z^A-\sigma_z^B)^2,\rho\}$ and $\mathcal{L}_2\rho:=(\sigma_z^A+\sigma_z^B)\rho(\sigma_z^A+\sigma_z^B)
-\frac{1}{2}\{(\sigma_z^A+\sigma_z^B)^2,\rho\}$. Here, superscripts $A$ and $B$ refer to the first and second qubit, respectively, and $\gamma_1(t)$ and $\gamma_2(t)$ are decay rates depending on the details of the microscopic model. The master equation described by Eq. (\ref{example-2}) fulfills Eq. (\ref{comm-condition}). Yet, in general, the attractive steady-state manifold may be one-dimensional decoherence-free subspaces. Nevertheless, by handling the spatial positions of the qubits such that they are very close with respect to the bath coherence length, the system can be controlled to experience collective decoherence. As is well-known, the non-trivial decoherence-free subspace can then appear in the system. This example can be generalized to the pure decoherence model of arbitrary qubits \cite{Palma1996,Reina2002,Chruscinski2010}.

In conclusion, we have found a family of non-Markovian dissipative processes that possess the same positive features as Markovian dissipative processes. Specifically, we have proved that a steady-state manifold always exists in the non-Markovian dissipative processes described by the general master equation (\ref{TCL-eqn}) with $\mathcal{L}(t)$ satisfying the commutation $[\mathcal{L}(t),\mathcal{L}(t^\prime)]=0$.  The structure of the steady-state manifold is expressed by Eq. (\ref{stru.}), which is the same as that in the Markovian dissipative processes, and the manifold is attractive if the attractive condition (\ref{conditon-for-attractiveness}) is satisfied. Our finding provides a promising way for implementing dissipation-assisted schemes in quantum information processing. By taking advantage of the same positive features, i.e., the existence, the structure, and the attractiveness of the steady-state manifold, the dissipation-assisted schemes implemented in the Markovian systems can be directly generalized to the non-Markovian systems.

This work was supported by NSF China through Grant No. 11575101. H. L. H. acknowledges support from NSF China through Grant No. 11571199. D. M. T. acknowledges support from the National Basic Research Program of China through Grant No. 2015CB921004.

\end{document}